\documentclass{iaus}
\usepackage[utf8]{inputenc}
\usepackage[english]{babel}
\usepackage{amssymb}
\usepackage{graphicx}

\newcommand{\aap}{{\it A\&A}}

\newcommand{\aj}{{\it AJ}}
\newcommand{\apj}{{\it ApJ}}

\newcommand{\mnras}{{\it MNRAS}}

\def\mstar{\hbox{$M_{\star}$}}
\def\msun{\hbox{${\rm M}_{\odot}$}}
\def\vsini{\hbox{$v\sin i$}}
\def\Prot{\hbox{$P_{\rm rot}$}}

\def\mrpd{\hbox{mrad\,d$^{-1}$}}
\def\ea{et al. }
\def\ie{i.e. }
\def\eg{e.g., }

\title[Magnetic topologies of cool stars] %
{Exploring the magnetic topologies\\ of cool stars}

\author[J. Morin, J.-F. Donati, P. Petit \ea]%
  {J. Morin$^{1,2}$, J.-F.~Donati$^2$, P.~Petit$^2$, %
   L.~Albert$^3$, M.~Aurière$^2$, R.~Cabanac$^2$, C.~Catala$^4$, %
   X.~Delfosse$^5$, B.~Dintrans$^2$, R.~Fares$^2$, T.~Forveille$^5$, %
   T.~Gastine$^2$, M.~Jardine$^7$, R.~Konstantinova-Antova$^6$, J.~Lanoux$^8$, %
   F.~Lignières$^2$, A.~Morgenthaler$^2$, F.~Paletou$^2$, %
   J.C.~Ramirez Velez$^4$, S.K.~Solanki$^9$, S.~Théado$^2$, V.~Van Grootel$^2$}

\affiliation{%
  $^1$Dublin Institute for Advanced Studies, 31 Fitzwilliam Place, %
    Dublin 2, Ireland\\%
    email: {\tt jmorin@cp.dias.ie} \\[\affilskip]%
  $^2$LATT, Université de Toulouse, CNRS, 14 Av. E. Belin, %
    31400 Toulouse, France\\[\affilskip]%
  $^3$CFHT, 65-1238 Mamalahoa Hwy, Kamuela HI 96743, USA\\[\affilskip]%
  $^4$LESIA, Observatoire de Paris-Meudon, 92195 Meudon, France\\[\affilskip]%
  $^5$LAOG, UMR5571 CNRS, Université Joseph Fourier, %
    BP 53, 38041 Grenoble, France\\[\affilskip]%
  $^6$Institute of Astronomy, Bulgarian Academy of Sciences, %
  72 Tsarigradsko shose, Sofia, Bulgaria\\[\affilskip]%
  $^7$School of Physics and Astronomy, University of St Andrews, %
      St Andrews, Scotland KY16 9SS\\[\affilskip]%
  $^8$Centre d’Etude Spatiale des Rayonnements, Université de Toulouse, %
  CNRS, France\\[\affilskip]%
  $^9$Max-Planck Institut für Sonnensystemforschung, Katlenburg-Lindau, %
  Germany\\[\affilskip]%
}

\pubyear{2011}
\volume{273}
\pagerange{1--6}
 \date{xxx and in revised form xxx}
\jname{Physics of Sun and Star Spots}
\editors{D.P.\ Choudhary \& K.G.\ Strassmeier, eds.}

\begin{document}

\maketitle

\begin{abstract}
Magnetic fields of cool stars can be directly investigated through the study of
the Zeeman effect on photospheric spectral lines using several approaches. With
spectroscopic measurement in unpolarised light, the total magnetic flux
averaged over the stellar disc can be derived but very little information on
the field geometry is available. Spectropolarimetry provides a complementary
information on the large-scale magnetic topology. With Zeeman-Doppler Imaging
(ZDI), this information can be retrieved to produce a map of the vector
magnetic field at the surface of the star, and in particular to assess the
relative importance of the poloidal and toroidal components as well as the
degree of axisymmetry of the field distribution.

The development of high-performance spectropolarimeters associated with
multi-lines techniques and ZDI allows us to explore magnetic topologies
throughout the Hertzsprung-Russel diagram, on stars spanning a wide range of mass, age
and rotation period. These observations bring novel constraints on magnetic
field generation by dynamo effect in cool stars. In particular, the study of
solar twins brings new insight on the impact of rotation on the solar dynamo,
whereas the detection of strong and stable dipolar magnetic fields on
fully convective stars questions the precise role of the tachocline in this
process.


\end{abstract}

\firstsection
\section{Context: stellar dynamos}
Magnetic field is a key parameter to understand stellar formation and
evolution.  In cool stars, it powers activity phenomena that are observed
across a large part of the electromagnetic spectrum and a wide range of
timescales. Since the early \textsc{XX}$^{\rm th}$ century, the cyclic solar
magnetic field has been thought to be constantly regenerated against ohmic
dissipation by a magnetohydrodynamical process: the dynamo. Although the solar
dynamo is still far from being fully understood,  the basic concepts, as
exposed by \cite{Parker55} in his $\alpha\Omega$ dynamo model are rather
simple: (i) an initially poloidal magnetic field is converted into a stronger
toroidal one by differential rotation, (ii) a poloidal field component is
regenerated from the toroidal field by a second mechanism, such as the $\alpha$
effect.  Two decades ago, helioseismology revealed the existence of the
tachocline (\eg \cite[Spiegel \& Zahn 1992]{Spiegel92}), a thin layer of strong
shear located at the interface between the inner radiative zone and the
convective envelope. Since then, many theoretical and numerical studies have
stressed the crucial role of the tachocline in the solar dynamo, being the
place where large-scale toroidal fields can be stored and strongly amplified
(\eg \cite[Charbonneau \& MacGregor 1997]{Charbonneau97}).

Partly convective cool stars possess an internal structure similar to that of
the Sun, \ie an inner radiative zone and an outer convective envelope
supposedly separated by a tachocline. Hence, it is generally assumed that their
magnetic fields --- as revealed by activity or direct measurements --- are
generated by a solar-like dynamo. However, some cool partly-convective stars
strongly differ from the Sun, either in depth of their convective zone or
rotation rate, and the impact of these differences on their dynamo is mostly
unknown.  On the other hand, main sequence stars less massive than $\sim
0.35~\msun$ ($\sim$M3) are fully convective (\eg \cite[Chabrier \& Baraffe
1997]{Chabrier97}) and therefore do not possess a tachocline. If the tachocline
is indeed an essential part of the solar dynamo, magnetic field generation in
these fully convective objects must rely on different physical processes.

\section{Magnetic field measurement and modelling}
Direct measurements of stellar magnetic fields rely on the properties of the
Zeeman effect. Two complementary methods are successfully  applied to cool
stars. By measuring Zeeman broadening of photospheric spectral lines it is
possible to assess the magnetic flux averaged over the visible stellar disc
(\eg \cite[Saar 1988]{Saar88}). This method is therefore able to probe magnetic
fields regardless of their complexity but provides virtually no information
about the field geometry. On its part, the analysis of Zeeman polarisation in
spectral lines provides information on the vector properties (\ie strength,
orientation and polarity). However, as neighboring magnetic regions of opposite
polarities result in polarised signatures of opposite sign that cancel each
other when integrating over the stellar disc, spectropolarimetry can only
detect the large-scale component of stellar magnetic fields.

Polarized signatures in spectral lines of cools active stars have a small
amplitude, making their measurement a hard task. Two advances have brought a
large number of cool stars within reach of spectropolarimetric measurements.
First, multi-line techniques (such as LSD, \cite[Donati \ea 1997]{Donati97a})
extract the polarimetric information from a large number of photospheric lines
resulting in a S/N multiplex gain that can reach as high as several tens when
thousands of lines are used. Secondly, the new generation spectropolarimeters
ESPaDOnS and NARVAL (see \cite[Donati 2003]{Donati03b}) feature a high overall
efficiency and cover the full optical domain allowing to take full advantage of
multi-line techniques.

Thanks to (\textit{i}) the sensitivity of the Zeeman effect to field lines
orientation (\textit{ii}) rotational modulation and (\textit{iii}) Doppler
effect, the temporal evolution of polarised signatures in stellar lines
strongly characterizes the parent magnetic topology. Thus, from a time-series
of circularly polarised spectra sampling the stellar rotation cycle,
Zeeman-Doppler Imaging can perform a maximum entropy reconstruction of the
vector magnetic field distribution at the surface of the star (\cite[Semel
1989]{Semel89}, \cite[Donati \& Brown 1997]{Donati97b}). Although the resulting
magnetic map has a better resolution for high \vsini\ values, this technique is
also successfully applied to slow rotators (\eg \cite[Morin \ea
2008b]{Morin08b}). For high S/N data spanning several stellar rotations,
differential rotation can also be constrained (\eg \cite[Petit \ea
2002]{Petit02}).

\section{Solar twins}
We selected a sample of 4 nearby dwarfs with stellar parameters as close as
possible to the solar ones, except for the rotation period (see \cite[Petit \ea
2008]{Petit08}). In particular, their internal structure is expected to be very
similar to the Sun's. We aim at studying the effect of rotation alone on the
solar dynamo. For each star, a set of $\sim$ 10 pairs of unpolarised and
circularly polarised spectra was collected with NARVAL, from which we can map
the surface magnetic field and to precisely determine the rotation period.

\begin{figure}
\label{fig:soltwin_pol}
\begin{center}
  \includegraphics[width=0.75\textwidth]{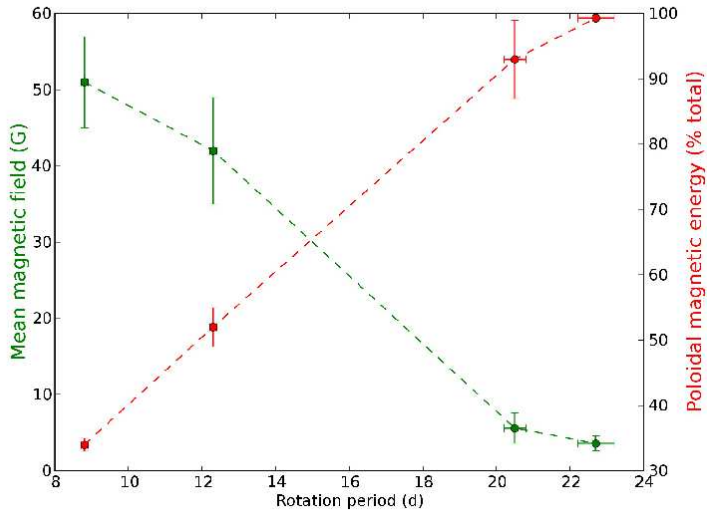}%
  \caption{Rotational dependence of the mean reconstructed magnetic flux (green
  line), and of the fraction of magnetic energy stored in the poloidal field
  component (red line).}
\end{center}
\end{figure}

The first conclusion of this study is that the shorter the rotation period, the
stronger is the large-scale magnetic field (see Fig.~1, green line). For the 2
slow rotators (\Prot=22.7 and 20.5~d), the magnetic maps reconstructed by ZDI
are dominated by low order multipole modes, this is reminiscent of the solar
global magnetic field. The fast rotators (\Prot=12.3 and 8.8~d) mainly feature
a strong belt of toroidal field roughly encircling the pole, similar to the
magnetic topologies of very active cool stars (\eg \cite[Donati \ea
2003]{Donati03a}). The transition from an almost purely poloidal magnetic field
to a strongly toroidal one, is apparently due to rotation, with a \Prot\
threshold located between 12 and 20~d (see Fig.~1). These observations are in
agreement with recent MHD simulations of solar-type stars where dynamo action
produces mostly toroidal magnetic topologies, featuring strong belts throughout
the bulk of the convective envelope for $\Omega=3~\Omega_\odot$ (\cite[Brown
\ea 2010]{Brown10}, although the simulation domain does not encompass the
stellar surface).

From our unpolarised spectra we measure the Ca \textsc{ii} activity, and find
that the $R'_{HK}(B_{mean})$ relation follows a power-law with an exponent of
0.32. This is significantly different from the solar relation (established from
observations of the quiet Sun and active regions) which has an exponent of 0.6
(\eg \cite[Schrijver et al. 1989]{Schrijver89}).  As the chromospheric flux is
also sensitive to spatial scales smaller than those contributing to the
polarimetric signal, this apparent discrepancy suggests that a larger fraction
of the total magnetic energy lies in the large-scale component as rotation
increases.

The stars we have observed are expected to exhibit magnetic cycles and
associated evolution of their topology. Chromospheric activity monitoring
exists for two stars of our sample (\eg \cite[Hall \ea 2007]{Hall07}), showing in
particular that our slowest rotator (\Prot=22.7~d) undergoes an activity cycle
of 7~yr, and that we observed it in a high activity state. The predominantly
quadrupolar magnetic topology we reconstruct is indeed reminiscent of the Sun's
topology close to solar maximum (\cite[Sanderson et al. 2003]{Sanderson03}). On
the fastest rotator (\Prot=8.8~d) we observe strong year-to-year evolution:
between 2007 and 2008 the polarity of the main ring of azimuthal field had its
polarity reversed, and between 2008 and 2009 the fraction of magnetic energy
stored in poloidal field dramatically increased (see \cite[Petit \ea
2009]{Petit09}). These rapid changes suggest a short cycle, and therefore a
correlation: faster rotation would imply shorter activity cycles.
Spectropolarimetric observations of the rapid rotator $\tau$ Boo (\Prot=3.3~d)
by \cite{Fares09} have also revealed two polarity reversals of the poloidal
field component in two years, although in this case the higher stellar mass
(\mstar=1.3~\msun) and the close-in orbiting giant planet may also play a role.

\section{Fully convective stars}
Following the first detection in polarised light of a large-scale magnetic
field on a fully convective star by \cite{Donati06}, we have carried out the
first spectropolarimetric survey of a sample M~dwarfs lying on both sides of
the fully convective boundary ($0.08<\mstar<0.75~\msun$) and spanning a wide
range of periods ($0.33<\Prot<18.6$~d). A total of 23 stars were selected, all
are active so that we can supposedly detect polarised signatures and map their
surface magnetic field. For each star we collected one or more time-series of
unpolarised and circularly polarised spectra. We could reconstruct a map of the
large-scale surface magnetic field and derive an accurate period measurement of
18 stars.  For the 5 remaining stars some constraints about the magnetic
topology and an upper limit for the rotation period could still be derived. For
more details see \cite{Donati08} and \cite[Morin \ea (2008a,b, 2010)]{Morin08a,
Morin08b, Morin10}.

\begin{figure}
\label{fig:mdw_plotMP}
\begin{center}

  \includegraphics[width=0.85\textwidth]{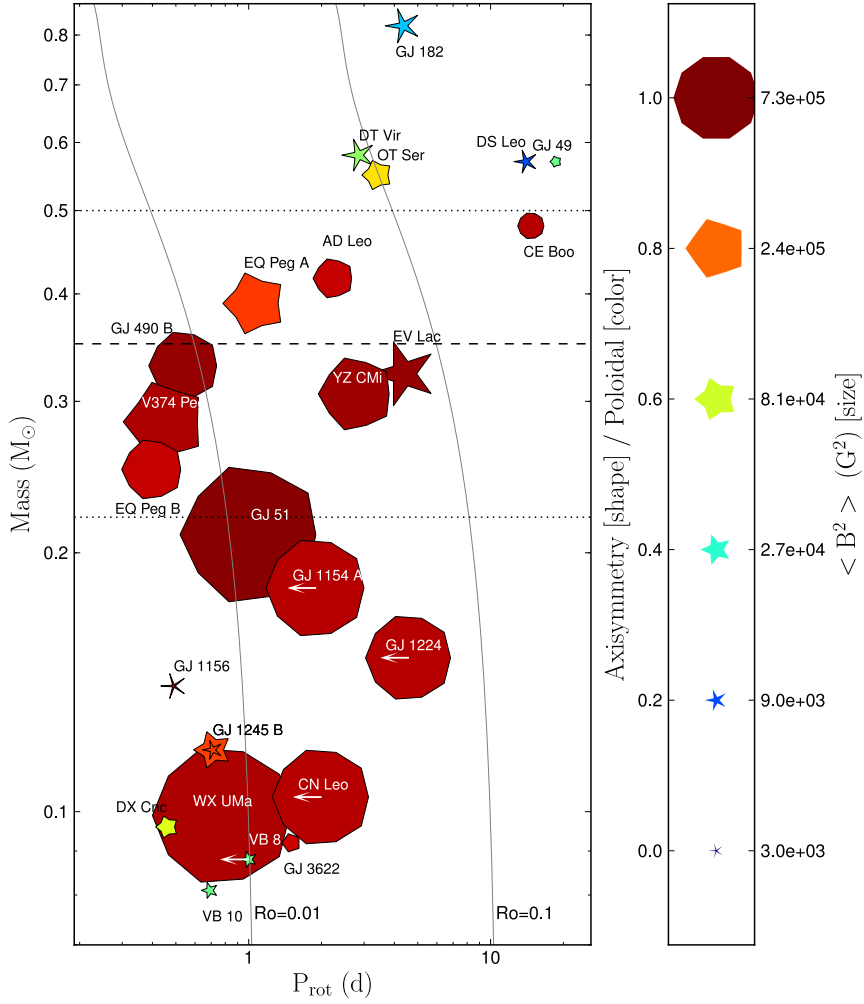}%
  \caption{Properties of the magnetic topologies of our sample of M dwarfs
  (plus GJ~490~B, \cite[Phan-Bao \ea 2009]{PhanBao09}) as a function of
  rotation period and mass. Larger symbols indicate stronger fields, symbol
  shapes depict the degree of axisymmetry of the reconstructed magnetic field
  (from decagons for purely axisymmetric to sharp stars for purely non
  axisymmetric), and colours the field configuration (from blue for purely
  toroidal to red for purely poloidal).  Solid lines represent contours
  of constant Rossby number $Ro=0.1$ (saturation threshold) and $0.01$. The
  theoretical full-convection limit ($\mstar \sim0.35~\msun$) is plotted as a
  horizontal dashed line, and the approximate limits of the three stellar
  groups discussed in the text are represented as horizontal solid lines.}

\end{center}
\end{figure}

The main results of this study are presented on Fig.~2, presenting the main
properties of the reconstructed magnetic topologies as a function of stellar
mass and rotation period. Our analysis reveals 3 distinct groups in this
diagram. 

\begin{enumerate}
\item M dwarfs more massive than $\sim0.5~\msun$ (partly convective) exhibit
magnetic topologies with a strong toroidal component, even dominant in some
cases; the poloidal component is strongly non-axisymmetric. For most of these
stars, we can measure surface differential rotation, values are between 60 and
120~\mrpd\ (\ie between once and twice the solar rate approximately), and the
topologies evolve beyond recognition on a timescale of a few months. These
properties are reminiscent of the observations of more massive (G and K) active
stars (\eg \cite[Donati \ea 2003]{Donati03a}).
\item Stars with masses between $\sim$0.2 and 0.5~\msun\ (close the fully
convective limit) host much stronger large-scale magnetic field with radically
different topologies: almost purely poloidal, generally nearly axisymmetric,
always close to a dipole more or less tilted with respect to the rotation axis.
These topologies are observed to be stable on timescales of several years, and
differential rotation (when measurable) is of the order or a tenth of the solar
rate.  Our findings are in partial agreement with the recent numerical study by
\cite{Browning08}. Similarly, we observe that fully convective stars can
generate strong and long-lived large-scale magnetic fields featuring a strong
axisymmetric component, that are able to quench differential rotation. But we
observe almost purely poloidal surface magnetic fields, whereas in the
simulation the axisymmetric component of the field is mainly toroidal (although
the simulation does not encompass the stellar surface). 
\item Below $\sim$0.2~\msun, we observe 2 different categories of magnetic
fields: either a very strong dipole (similar to group \textit{b} above) or a
much weaker field generally featuring a significant non-axisymmetric component,
and in some cases a toroidal one. Strong temporal variability is also observed
on some objects of this second category. However stars in both categories have
similar stellar parameters and cannot be separated in the mass-rotation
diagram. This unexpected observation is not yet understood, and may be
explained in several different ways: for instance, another parameter than mass
and rotation period (such as age) may be relevant, two dynamo modes may be
possible or stars may switch between two states in this mass range, etc.
\end{enumerate}

\section{Conclusions and future work}
Taking advantage of recent developments in spectropolarimetric instrumentation
and analysis techniques, it is nowadays possible to study dynamo action across
the whole cool stars regime. We focus here on two projects: solar twins and
fully convective M dwarfs. Both studies have already provided dynamo theorists
with novel observational constraints, and have also raised new questions on
the impact of magnetic fields and their topology on \eg stellar spindown,
structure, and chromospheric and coronal activity.

Our understanding of stellar dynamos also benefits from spectropolarimetric
studies targeting other objects than cool main sequence stars. For instance,
the first results obtained on a few T Tauri stars located on both sides of the
fully convective boundary reveal similarities with low-mass main
sequence stars (\eg \cite[Donati et al. 2010]{Donati10}), while the
spectropolarimetric discovery of a magnetic field on Betelgeuse suggests that
very-slowly-rotating supergiants can sustain local dynamo action (\cite[Aurière
et al. 2010]{Auriere10}).

Ongoing and future work includes in particular long-term monitoring of a small
sample of selected targets to study magnetic cycles, and the extension of the
survey to cover the mass-rotation plane on the whole cool stars regime.

\begin{acknowledgments}
The presentation of this paper in the IAU Symposium 273 was possible due to
partial support  from the  National Science Foundation grant numbers ATM
0548260, AST 0968672 and NASA - Living With a Star grant number
09-LWSTRT09-0039.
\end{acknowledgments}


\begin{thebibliography}{}

\bibitem[Aurière et al. (2010)]{Auriere10}
  Aurière M., \ea, 2010, \aap, 516, L2 

\bibitem[Brown et al. (2010)]{Brown10} 
  Brown B.~P., Browning M.~K., Brun A.~S., Miesch M.~S., Toomre J.,
  2010, \apj, 711, 424 

\bibitem[Browning (2008)]{Browning08} 
  Browning M.~K., 2008, \apj, 676, 1262 

\bibitem[Chabrier \& Baraffe (1997)]{Chabrier97}
  {Chabrier} G.,  {Baraffe} I.,  1997, \aap, 327, 1039

\bibitem[Charbonneau \& MacGregor (1997)]{Charbonneau97} 
  Charbonneau P., MacGregor K.~B., 1997, \apj, 486, 502 

\bibitem[Donati \ea (1997)]{Donati97a}
  {Donati} J.-F.,  {Semel} M.,  {Carter} B.~D.,  {Rees} D.~E.,
  {Cameron} A.~C.,  1997, \mnras, 291, 658

\bibitem[Donati \& Brown (1997)]{Donati97b}
  {Donati} J.-F.,  {Brown} S.~F.,  1997, \aap, 326, 1135

\bibitem[Donati \ea (2003)]{Donati03a} 
Donati J.-F., et al., 2003, \mnras, 345, 1145 

\bibitem[Donati (2003)]{Donati03b}
  Donati J.-F., 2003, \textit{ASPC}, 307, 41 

\bibitem[Donati \ea (2006)]{Donati06}
  {Donati} J.-F., \ea,  2006, \textit{Science}, 311, 633

\bibitem[Donati \ea (2008)]{Donati08}
  {Donati} J.-F., \ea, \mnras, 390, 545

\bibitem[Donati et al. (2010)]{Donati10} 
  Donati J.-F., et al., 2010, \mnras, 402, 1426 

\bibitem[Fares \ea (2009)]{Fares09} 
  Fares R., et al., 2009, \mnras, 398, 1383 

\bibitem[Hall \ea (2007)]{Hall07} 
  Hall J.~C., Lockwood G.~W., Skiff B.~A., 2007, \aj, 133, 862 


\bibitem[Morin \ea (2008a)]{Morin08a}
  {Morin} J., \ea,  2008a, \mnras, 384, 77

\bibitem[Morin \ea (2008b)]{Morin08b}
  {Morin} J., \ea,  2008b, \mnras, 390, 567

\bibitem[Morin \ea (2010)]{Morin10} 
  Morin J., Donati J.-F., Petit P., Delfosse X., Forveille T., Jardine M.~M.,
  2010, \mnras, 1077 

\bibitem[Parker (1955)]{Parker55}
  {Parker} E.~N.,  1955, \apj, 122, 293

\bibitem[Petit \ea (2002)]{Petit02}
  Petit P., Donati J.-F., Collier Cameron A., 2002, \mnras, 334, 374 

\bibitem[Petit \ea (2008)]{Petit08} 
  Petit P., et al., 2008, \mnras, 388, 80 

\bibitem[Petit \ea (2009)]{Petit09} 
  Petit P., \ea, 2009, \aap, 508, L9 

\bibitem[Phan-Bao \ea (2009)]{PhanBao09}
  Phan-Bao N., Lim J., Donati J.-F., Johns-Krull C.~M., Mart{\'{\i}}n E.~L.,
  2009, \apj, 704, 1721 

\bibitem[Saar (1988)]{Saar88}
  Saar S.~H., 1988, \apj, 324, 441 

\bibitem[Sanderson et al.(2003)]{Sanderson03}
  Sanderson T.~R., Appourchaux T., Hoeksema J.~T., Harvey K.~L.,
  2003, \textit{JGRA}, 108, 1035 

\bibitem[Schrijver et al. (1989)]{Schrijver89}
  Schrijver C.~J., Cote J., Zwaan C., Saar S.~H., 1989, \apj, 337, 964 

\bibitem[Semel (1989)]{Semel89}
  Semel M., 1989, \aap, 225, 456

\bibitem[Spiegel \& Zahn (1992)]{Spiegel92}
  {Spiegel} E.~A.,  {Zahn} J.-P.,  1992, \aap, 265, 106

\end{thebibliography}
\end{document}